# Ultra-low-loss $Ta_2O_5$ waveguide for integrated nonlinear photonics

XIAOTONG ZHANG,[1,†] SHUXUAN GUO,[2,3,†] RUIXUAN YI,[1,†] ZHEZE SHI,[1] LIKE SHUI,[4] LONG HUANG,[2,3] LEIRAN WANG,[2,3] YANG WANG,[2,3,*] JUNLING QU,[1,*] WENFU ZHANG,[2,3] XUETAO GAN[1,4,*]

[1]*Key Laboratory of Light Field Manipulation and Information Acquisition, Ministry of Industry and Information Technology, and Shaanxi Key Laboratory of Optical Information Technology, School of Physical Science and Technology, Northwestern Polytechnical University, Xi'an 710129, China*
[2]*State Key Laboratory of Ultrafast Optical Science and Technology, Xi'an Institute of Optics and Precision Mechanics, Chinese Academy of Sciences, Xi'an 710119, China*
[3]*University of Chinese Academy of Sciences, Beijing 100049, China*
[4]*School of integrated circuits and Microelectronics, Northwestern Polytechnical University, Xi'an 710129, China*
[†]*These authors contributed equally.*
** wangyang2017@opt.ac.cn, jlqu@nwpu.edu.cn, xuetaogan@nwpu.edu.cn*

**Abstract**: Integrated nonlinear photonics demands waveguides that simultaneously offer ultra-low-loss, relatively strong nonlinearity, and tight mode confinement. Tantalum pentoxide ($Ta_2O_5$) is attractive for this purpose because it combines a broad transparency window, relatively strong third-order nonlinearity, and favorable thick-film growth characteristics. In tightly confined $Ta_2O_5$ waveguides, however, enhanced interaction between the optical mode and etched sidewalls makes scattering loss an important limitation for achieving high $Q$ anomalous-dispersion resonators. Here, we mitigate this confinement-scattering trade-off in pure $Ta_2O_5$ waveguides using a fabrication flow based on the Damascene concept. The resulting microring resonators (MRRs) with a cross-section of 1.6 μm × 0.85 μm exhibit intrinsic quality factors exceeding $4.0 \times 10^6$, corresponding to propagation losses below 10 dB/m, while maintaining anomalous dispersion suitable for bright dissipative Kerr soliton operation. Using these resonators, we generate a coherent bright single-soliton microcomb with a repetition rate of 108.32 GHz. Under optimized dual-pump operating conditions, the platform further exhibits an experimentally accessible single-soliton window spanning 7.5 GHz, indicating a large detuning tolerance for soliton operation; this broad operating window is consistent with the relatively weak thermo-optic response of $Ta_2O_5$. Thermal-bistability analysis further indicates that the residual scattering loss is comparable to the material-absorption loss, rather than dominating the loss budget. These results demonstrate that low optical loss, tight mode confinement, and dispersion engineering suitable for soliton formation can be jointly achieved in a pure $Ta_2O_5$ platform, providing an ultra-low-loss platform foundation for integrated nonlinear photonic devices and more complex chip-scale nonlinear photonic systems.

## 1. Introduction

Integrated nonlinear photonics provides an important foundation for chip-scale ultrafast all-optical signal generation and processing, with applications spanning optical communications, optical computing, and quantum information [1–3]. On-chip sources based on third-order nonlinearities, including Kerr frequency combs [4] and supercontinuum generation [5], can provide broad bandwidth and high coherence, supporting applications ranging from optical communications to precision metrology and high-sensitivity sensing [6]. Efficient nonlinear optical processes generally require waveguides or resonators that simultaneously provide low optical loss and tight mode confinement. Low propagation loss extends the effective light–matter interaction length and enhances intracavity optical fields, whereas tight mode confinement reduces the effective mode area, enhances the effective nonlinearity, and provides increased flexibility for dispersion engineering.

Integrated nonlinear photonics has been widely investigated on material platforms including silicon-on-insulator (SOI), lithium-niobate-on-insulator (LNOI), silicon nitride ($Si_3N_4$), and chalcogenide glasses [7]. These platforms offer different combinations of optical nonlinearity, propagation loss, dispersion control, and fabrication characteristics. SOI suffers from two-photon absorption in the telecommunication band [8], while the fabrication and etching of LNOI remain challenging [9], and chalcogenide glasses can exhibit limited thermal and process compatibility [10]. Among these platforms, $Si_3N_4$ has been extensively developed for Kerr nonlinear photonics. Stoichiometric $Si_3N_4$ waveguides fabricated by low-pressure chemical vapor deposition (LPCVD) can achieve extremely low propagation loss [11,12], although thick films can suffer from tensile-stress-induced cracking and therefore require stress-mitigation strategies [13]. Low-temperature plasma-enhanced chemical vapor deposition (PECVD) enables thick $Si_3N_4$ films with tight mode confinement, but N–H-related absorption near the telecommunication band can increase propagation loss [14].

$Ta_2O_5$ has become an established material platform for integrated Kerr nonlinear photonics. It combines a broad optical transparency window and a refractive index of approximately 2 with a relatively strong third-order nonlinear response. Previous studies have characterized its Kerr nonlinear refractive index, thermo-optic response, and annealing behavior, and have demonstrated low-loss waveguides, high *Q* microresonators, and dissipative Kerr soliton (DKS) [15–18]. These studies established an important material and device foundation for $Ta_2O_5$ Kerr nonlinear photonics.

More recently, $Ta_2O_5$ and related tantalum-oxide photonic platforms have been extended in several directions. MacFarlane et al. explored magnetron-sputtered $NbTaO_x$ for integrated linear and nonlinear photonics, including optical parametric oscillation and supercontinuum generation [19]. Brodnik et al. demonstrated monolithic three-dimensional integration of $Ta_2O_5$ with thin-film lithium niobate, combining $\chi^{(3)}$ and $\chi^{(2)}$ nonlinear functionalities [20]. Carollo et al. employed composition engineering in amorphous metal-oxide mixtures to achieve high *Q* nonlinear photonic resonators [21], while Zang et al. demonstrated highly efficient soliton formation in a bidirectional Kerr resonator [22]. In addition, Lucas et al. introduced inverse-designed meta-dispersion microresonators for advanced microcomb control [23]. Together, these studies illustrate the progression of tantalum-oxide nonlinear photonics from fundamental material and device demonstrations toward high *Q* resonators, efficient microcomb operation, advanced dispersion control, and multifunctional integration.

Despite these advances, simultaneously maintaining low optical loss and tight mode confinement remains an important device challenge in conventional geometry-engineered pure $Ta_2O_5$ microresonators. Tight mode confinement is desirable for enhancing the effective nonlinearity and enabling geometric dispersion engineering, which can be used to tailor the phase-matching conditions for Kerr four-wave mixing and to realize the anomalous dispersion required for conventional bright-DKS formation. At the same time, tighter confinement increases the interaction of the optical mode with the core–cladding interfaces and consequently enhances sensitivity to etched-sidewall roughness and scattering loss [24,25]. Previous work has demonstrated propagation losses as low as approximately 3 dB/m using an ultrathin, weakly confined $Ta_2O_5$ core [26]. However, such weakly confined geometries are less suitable for the tightly confined devices considered here, where dispersion is engineered through the waveguide geometry. Jung et al. demonstrated bright-DKSs in anomalous-dispersion $Ta_2O_5$ microrings with waveguide widths of 1.6 and 1.65 μm [15]. The highest intrinsic quality factor ($Q_i$) reported in that study, ($Q_i \approx 3.8 \times 10^6$), was measured in a wider, approximately 5-μm-wide ring designed to reduce mode–sidewall interaction [15]. These observations are consistent with the device-level trade-off between tight mode confinement and scattering loss. Building on these advances, an important objective is to realize multi-million $Q_i$ directly in a tightly confined anomalous-dispersion geometry used for bright-DKS operation.

Here, we address this device-level challenge using a fabrication flow based on the Damascene concept to reduce scattering loss in tightly confined pure $Ta_2O_5$ devices [27]. Waveguide trenches are defined in $SiO_2$ using deep ultraviolet (DUV) lithography and inductively coupled plasma (ICP) etching, followed by thermal reflow to reduce etched-sidewall irregularities. Pure $Ta_2O_5$ is subsequently deposited by magnetron sputtering, and chemical-mechanical polishing (CMP) is used to reduce the top-surface roughness. Using this fabrication flow, we realize tightly confined $Ta_2O_5$ microrings with a 1.6 μm × 0.85 μm cross-section, intrinsic quality factors $Q_i$ exceeding $4.0 \times 10^6$, and propagation losses below 10 dB/m, while retaining anomalous dispersion suitable for bright-DKS operation. Thermal-bistability analysis further indicates comparable contributions from optical absorption and scattering to the remaining propagation loss. Using the same class of anomalous-dispersion resonators, we generate a coherent bright single-soliton microcomb with a repetition rate of 108.32 GHz, extending conventional anomalous-dispersion bright-DKS operation in $Ta_2O_5$ toward the ~100 GHz repetition-rate regime. Under optimized dual-pump operating conditions, an experimentally accessible single-soliton window spanning 7.5 GHz is further observed, indicating a large detuning tolerance for soliton operation. This broad operating window is consistent with the relatively weak thermo-optic response of $Ta_2O_5$. These results demonstrate that low propagation loss, tight mode confinement, anomalous dispersion, and practical Kerr-soliton operation can be jointly realized in a pure $Ta_2O_5$ platform through coordinated fabrication and dispersion engineering.

## 2. Results and Discussion

### *2.1 Device fabrication*

To fabricate an ultra-low-loss $Ta_2O_5$ photonic platform, we developed a specific fabrication workflow inspired by the Damascene process [27], as outlined in Figure 1(b). First, grooves of waveguide or microring resonator (MRR) patterns are defined in a 5-μm-thick $SiO_2$ layer thermally oxidized on silicon wafer with the combined processes of DUV lithography and ICP etching in a standard CMOS-foundry with 180-nm process. The waveguide grooves are defined with a width of 1.6 μm and depth of 0.85 μm to define the corresponding $Ta_2O_5$ core geometry. In particular, prior to $Ta_2O_5$ film deposition, a thermal reflow step of the waveguide grooves at 1050 °C for 5 hours in an oxygen atmosphere is performed. This step utilizes the surface-tension-driven reflow of $SiO_2$ at a temperature slightly above its glass transition temperature, which can mildly smooth the vertical striations on the dry-etched $SiO_2$ sidewalls while preserving the waveguide grooves geometry [28]. A quantitative root-mean-square (RMS) roughness value of the $SiO_2$ trench sidewall was not directly measured in the present work; therefore, the sidewall improvement is discussed here based on the observed morphological change in the scanning electron microscope (SEM) images. This smoothing effect can be evidenced by the top-view SEM images (see Supplementary Information Note 2), where the $SiO_2$ trench right after ICP etching shows a rough texture, while the one after thermal reflow is unambiguously smoother. Subsequently, the wafer is deposited with a $Ta_2O_5$ thin film using a magnetron sputtering system. This sputtering system is powered by a 50 W radio frequency source at a chamber pressure of 5 mTorr, with an argon to oxygen flow ratio of 1:0.25 to balance the deposition rate and film stoichiometry. The plasma bombards a $Ta_2O_5$ target (99.99% purity) and sputters it onto the waveguide grooves. To evaluate the thickness uniformity of the $Ta_2O_5$ sputtering process, a wafer was coated under the same deposition conditions. Thickness measurements at 17 predefined positions yielded a maximum thickness of ~854.9 nm and a minimum of ~850.5 nm, corresponding to a thickness variation of ± 2.2 nm (a non-uniformity of < 0.3%). In addition, full-wafer stress mapping yields an average residual stress of approximately −152.3 MPa, indicating an overall compressive stress state. The corresponding thickness and stress maps are provided in Supplementary Note 8. After $Ta_2O_5$ deposition with a target thickness, chemical mechanical polishing (CMP) is employed to planarize the sample surface. As revealed by the atomic force microscope (AFM) images (see Supplementary Information Note 3), the top surface roughness ($R_q$) of the waveguide decreases from 0.75 nm to 0.20 nm after the CMP step, which is expected to significantly mitigate scattering loss. Finally, thermal annealing is performed at 525 °C in an oxygen atmosphere to further reduce material absorption loss [29], as will be discussed later.

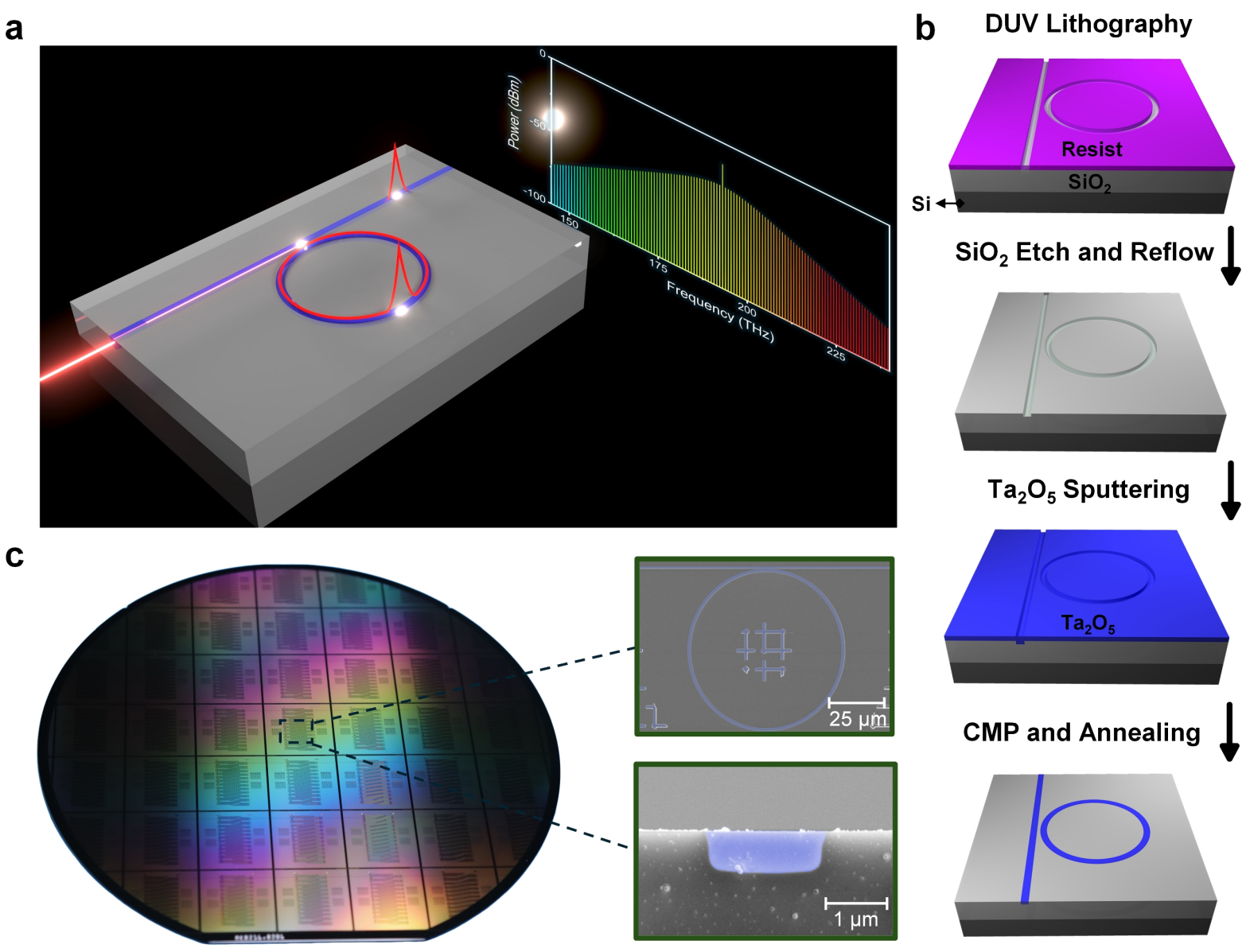

**Figure 1.** $Ta_2O_5$ nonlinear photonic platform and its fabrication. **(a)** Schematic of dissipative Kerr soliton generation in a $Ta_2O_5$ MRR driven by a continuous-wave laser, alongside the corresponding optical frequency comb spectrum. **(b)** Fabrication steps based on the Damascene process, including DUV lithography, $SiO_2$ etching and reflow, $Ta_2O_5$ sputtering, CMP and annealing. **(c)** Photograph of the fabricated

150-mm wafer, with inset SEM images showing a top-view of MRRs and a cross-section of the $Ta_2O_5$ waveguide. The final fabricated profiles are approximately 1.6 μm in width and 0.85 μm in height.

### *2.2 Quality factors of fabricated MRRs and propagation loss*

The propagation loss of the fabricated $Ta_2O_5$ waveguides is assessed by obtaining the $Q_i$ of the MRRs. The $Q_i$, representing the ratio of stored energy to the energy dissipated per optical cycle, is related to the waveguide propagation loss ($\alpha$) by [30]:

$$\alpha_{\mathrm{Np/m}} = \frac{2\pi n_{\mathrm{g}}}{Q_{\mathrm{i}}\lambda} \tag{1}$$

where $n_g$ is the group index of the waveguide mode and $\lambda$ is the vacuum wavelength. The group index was calculated using the Lumerical MODE solver based on the experimentally measured waveguide cross-section and the wavelength-dependent refractive index of the sputtered $Ta_2O_5$ film obtained by spectroscopic ellipsometry. Across the measured wavelength range of 1500 nm to 1600 nm, the calculated group index varies from approximately 2.179 to 2.183, with a specific value of 2.182 at 1550 nm. The corresponding propagation loss in dB/m is obtained as:

$$\alpha_{\mathrm{dB/m}} = 10\log(e)\alpha_{\mathrm{Np/m}} \tag{2}$$

For an unsplit resonance measured in the linear-response regime, the loaded quality factor is determined from the resonance linewidth, and $Q_i$ is related to $Q_L$ and the normalized minimum transmission $T_{min}$ by [31]:

$$Q_{\mathrm{i}} = \frac{2Q_{\mathrm{L}}}{1 \pm \sqrt{T_{\mathrm{min}}}} \tag{3}$$

where $T_{min}$ is the normalized transmission at the resonant wavelength, and the positive and negative signs correspond to the under-coupled and over-coupled regimes, respectively [32].

Because the coupling regime cannot be uniquely determined from a single transmission spectrum alone, we independently identified the coupling condition by characterizing a series of MRRs with systematically varied bus-waveguide-ring coupling gaps. With increasing gap, the normalized on-resonance transmission $T_{min}$ first decreases, reaches a minimum near critical coupling, and subsequently increases (Supplementary Note 4). A representative device on the larger-gap side of this transition was therefore identified as under-coupled and selected for the $Q_i$ statistics presented in Figure 2(e). For all selected resonances, the extracted quality factors satisfy $Q_c > Q_i$ and $Q_L^{-1} = Q_i^{-1} + Q_c^{-1}$.

All transmission spectra used for $Q$ factor extraction were measured in the low-input-power regime to minimize thermo-optic distortion. Resonances without resolvable mode splitting were fitted using a Lorentzian model, whereas resolvably split resonances were analyzed using a coupled-mode theory (CMT) model that accounts for clockwise and counterclockwise (CW-CCW) backscattering [33]. Resonances exhibiting pronounced thermo-optically induced asymmetric lineshapes were not used for the $Q_i$ and propagation-loss statistics. Detailed treatments of mode-split and thermo-optically distorted resonances, together with the fitting-uncertainty analysis, are provided in Supplementary Notes 5-7.

Based on the fabricated MRRs with a radius of 200 μm, we investigated the effect of the thermal annealing temperature on the waveguide propagation loss. As presented in Figure 2(a), for a constant annealing duration of 5 hours, the extracted $Q_i$ first increases with the annealing temperature, reaching approximately $4.0 \times 10^6$ at an annealing temperature of 525 °C. The increase in $Q_i$ with annealing temperature up to 525 °C is consistent with the suppression of oxygen-defect-related absorption. Oxygen vacancies and related sub-stoichiometric trap states have been widely reported in magnetron-sputtered $Ta_2O_5$ films, introducing additional optical absorption in the telecommunication band [29,34,35]. Although we did not perform X-ray photoelectron spectroscopy (XPS) to strictly quantify the atomic ratio of Ta and O, the significant improvement in $Q$ factor upon annealing provides indirect evidence consistent with the oxidation of sub-stoichiometric $TaO_x$ and the reduction of oxygen-vacancy-related absorption [15]. However, at higher annealing temperatures, crystallization becomes evident. Distinct crystalline peaks are observed at 600 °C and above, indicating the formation of crystalline grains and grain boundaries that can introduce significant scattering loss [36–38]. This crystallization of $Ta_2O_5$ was verified by the X-ray diffraction (XRD) patterns obtained at various temperatures, as shown in Figure 2(b). At temperatures below 500 °C, the $Ta_2O_5$ films are amorphous, characterized by a broad diffraction hump around 20° to 30°, alongside sharp peaks from the silicon substrate (marked with ♥). In contrast, upon annealing at 600 °C and above, distinct sharp peaks (marked with ♣) corresponding to the orthorhombic phase of $Ta_2O_5$ (β-$Ta_2O_5$, PDF card #25-0922) [39–41] emerge, confirming the phase transition from amorphous to crystalline.

With the optimal annealing temperature determined for the $Ta_2O_5$ film, we then examined the effect of the $SiO_2$ groove reflow process on the waveguide propagation loss. As indicated in Figures 2(c) and 2(d), the MRR device fabricated with $SiO_2$ reflow exhibits a significantly higher $Q_i$ of $(5.05 \pm 0.10) \times 10^6$, compared to its non-reflowed counterpart $(1.02 \pm 0.13) \times 10^6$, owing to the reduced scattering loss at the sidewalls. It is worth noting that while the non-reflowed device exhibits a broad, unsplit resonance, the significantly narrowed linewidth in the reflowed device makes the mode splitting induced by residual backscattering clearly resolvable [Fig. 2(d)]. To further evaluate the broadband performance, we present the statistical distributions of the quality factors in Figure 2(e). The results show the extracted intrinsic (red circles), loaded (blue circles), and coupling (green circles) quality factors within the wavelength range from 1500 nm to 1600 nm. Across the measured wavelength range, the extracted $Q_i$ values exhibit a mean of $(4.24 \pm 0.72) \times 10^6$, where the uncertainty denotes one standard deviation across the analyzed resonances. Importantly, no systematic degradation of $Q_i$ is observed around 1520 nm, indicating no obvious N–H-bond-related absorption feature in this spectral region. To further assess device-to-device consistency, we characterized multiple resonators across five different samples. The extracted $Q_i$ exhibit a relatively concentrated distribution with an average $Q_i$ of approximately $4.0 \times 10^6$, as summarized in Supplementary Note 8. Overall, these results validate the viability of our fabrication workflow for realizing a robust, ultra-low-loss $Ta_2O_5$ integrated photonic platform. Building upon these linear optical characterizations, we further exploited the thermal bistability effect of the MRRs to quantitatively disentangle the intrinsic material absorption and scattering losses from the total propagation loss.

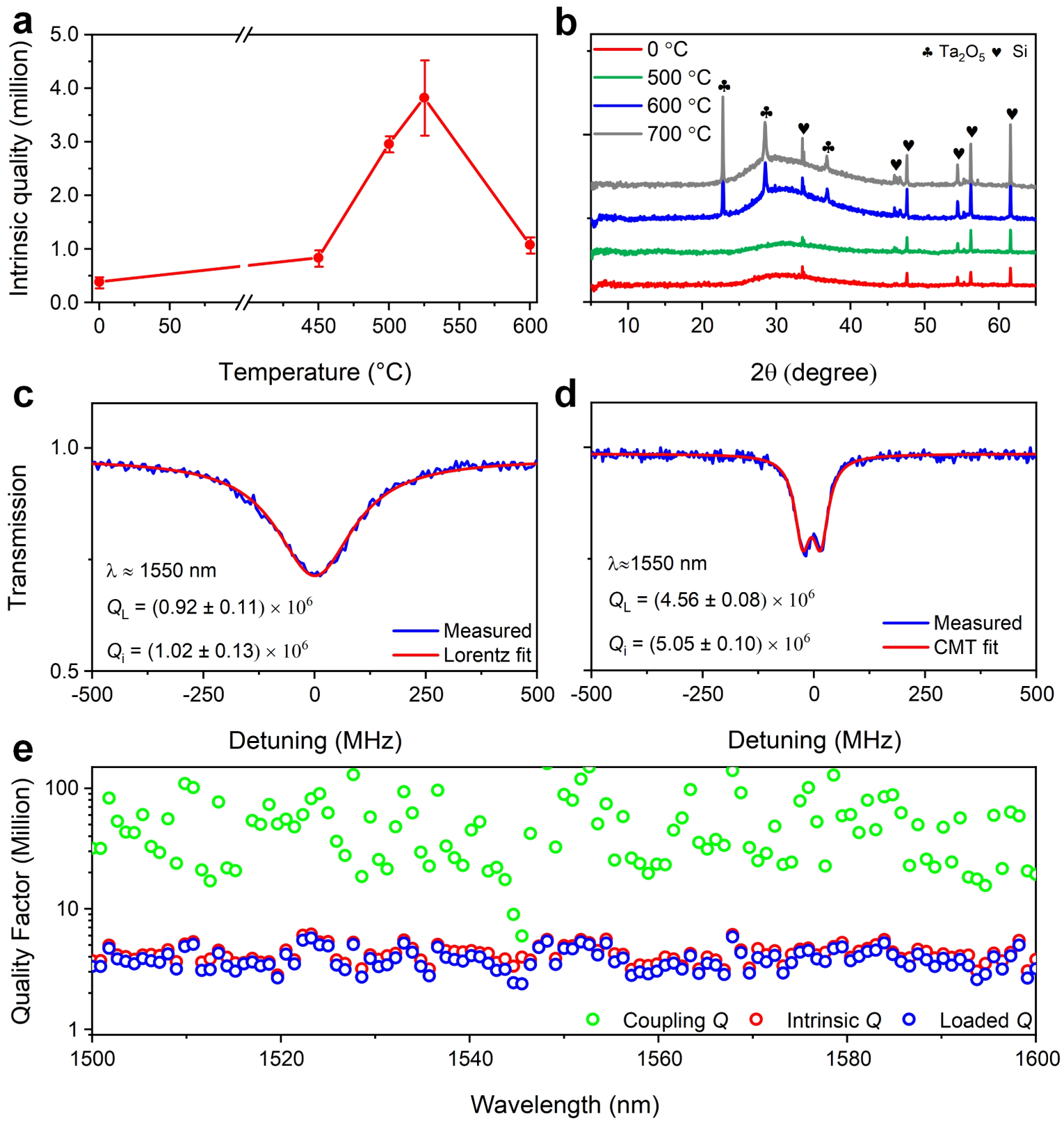

**Figure 2**. Fabrication optimizations and optical characterizations of $Ta_2O_5$ MRRs. **(a)** Extracted $Q_i$ as a function of the annealing temperature for a fixed annealing duration of 5 hours. **(b)** XRD patterns of the $Ta_2O_5$ films annealed at various temperatures. The symbols ♥ and ♣ denote the diffraction peaks of the silicon substrate and the crystallized orthorhombic $Ta_2O_5$ phase, respectively. **(c, d)** Normalized transmission spectra and corresponding Lorentz fits (c) and CMT fits (d) of the resonance modes for MRRs fabricated without and with the $SiO_2$ reflow process, respectively. The extracted $Q_L$ and $Q_i$, along with their fitting uncertainties, are indicated. (e) Statistical distributions of the intrinsic (red circles), loaded (blue circles), and coupling (green circles) quality factors measured over a broad wavelength range from 1500 nm to 1600 nm. The data in (e) are extracted from a representative under-coupled MRR ($Q_c > Q_i$).

### *2.3 Propagation loss analyzed by thermal bistability response*

Specifically, using the aforementioned $Ta_2O_5$ MRR, we conducted thermal bistability measurements [42]. During the characterization, a specific fraction (i.e., the absorption ratio, η) of the optical power coupled into the cavity is absorbed by the $Ta_2O_5$ waveguide material. This absorbed energy is converted into thermal energy, raising the cavity temperature and inducing a redshift of the resonance wavelength via the thermo-optic effect [17]. This frequency shift, in turn, alters the intracavity power, establishing a positive feedback loop that ultimately distorts the resonance lineshape from a symmetric Lorentzian to an asymmetric triangle [43].

Experimentally, we probed the transmission spectra of the MRR under on-chip input power ($P_{in}$) ranging from 0.3 mW to 0.7 mW. As shown in Figure 3(a), clear redshift and a triangular resonance lineshape centered near 1550 nm are observed. From these spectra, the resonance frequency shift $\delta\omega$ as a function of the dropped power $P_d$ is extracted in Figure 3(b), where $P_d = P_{in}\,(1\text{-}T_{res})$ represents the power coupled into the MRR at resonance. By defining a thermal response coefficient ($\chi_{th}$) as $\delta\omega = \chi_{th}\, P_d$ [28], we extracted a value of $\chi_{th}$ = 1048.91 MHz/mW from the linear fit between $\delta\omega$ and $P_d$. To determine the absorption ratio, we consider the steady-state thermal balance, where the absorbed power ($P_{\text{abs}}$) results in a temperature rise ($\Delta T$) governed by $P_{\text{abs}} = K_C\ \Delta T$. Here, $K_C$ represents the intrinsic thermal conductance of the waveguide system, characterizing its heat dissipation to the surroundings. Meanwhile, the frequency shift is proportional to the temperature change via $\delta\omega = \beta(\omega)\Delta T$, where $\beta(\omega)$ is the thermal frequency shift coefficient. By measuring the resonance frequency shift as a function of the chip temperature (controlled by a thermoelectric cooler stage), we obtained $\beta(\omega)$ = 791.63 MHz/K [Figure 3(c)]. By noting that the absorbed power is a fraction of the dropped power ($P_{\text{abs}} = \eta\ P_d$), the absorption ratio $\eta$ can be directly derived as $\eta = \chi_{th}\ K_C/\beta(\omega)$. Here, $K_C$ acts as a key fitting parameter extracted by modeling the device's transient thermal response. According to the thermodynamic model [44], the time-dependent effective temperature rise ($\Delta T_{eff}$) is governed by:

$$\Delta T_{eff} = \frac{P_{\text{abs}}}{K_C} + (T_0 - \frac{P_{\text{abs}}}{K_C})e^{-\gamma_T t} \tag{4}$$

where $t$ is the time, $\gamma_T$ is the thermal decay coefficient, and $T_{eff}$ ($\Delta T_{eff} = T_{eff} - 293.15$ K) is the effective temperature averaged over the optical mode volume [44]:

$$T_{eff} = \frac{\int T(r,t) n^2(r) \left| E(r) \right|^2 d^3 r}{\int n^2(r) \left| E(r) \right|^2 d^3 r} \tag{5}$$

where $n(r)$ is the position-dependent refractive index and $E(r)$ is the electric field of resonance mode. For simplicity, the spatial distribution of the heat source is assumed to follow the optical mode intensity distribution, as presented in the inset of Figure 3(d).

Using finite element method (FEM) simulations, we modeled the thermal dynamics of the MRR and obtained the transient traces of $\Delta T_{eff}$, shown as blue dots in Figure 3(d). By fitting the simulated data according to Eq. (4), plotted as a red line in Figure 3(d), the value of $K_C$ was extracted to be $4.078 \times 10^{-4}$ W/K. Consequently, this yields an $\eta$ of approximately 50%, indicating that material absorption and scattering losses contribute comparably to the total intrinsic loss.

To further improve the performance of this $Ta_2O_5$ platform, future optimizations can target both the absorption and scattering losses individually. Specifically, the absorption loss can be further mitigated by optimizing the oxygen-to-argon ratio during magnetron sputtering, or by adopting more advanced deposition techniques, such as ion beam sputtering (IBS) to achieve high-density, low-defect films. Regarding the sidewall scattering loss, the $SiO_2$ groove etching process can be further improved, for instance, by incorporating a photoresist reflow step to enhance the sidewall smoothness [45].

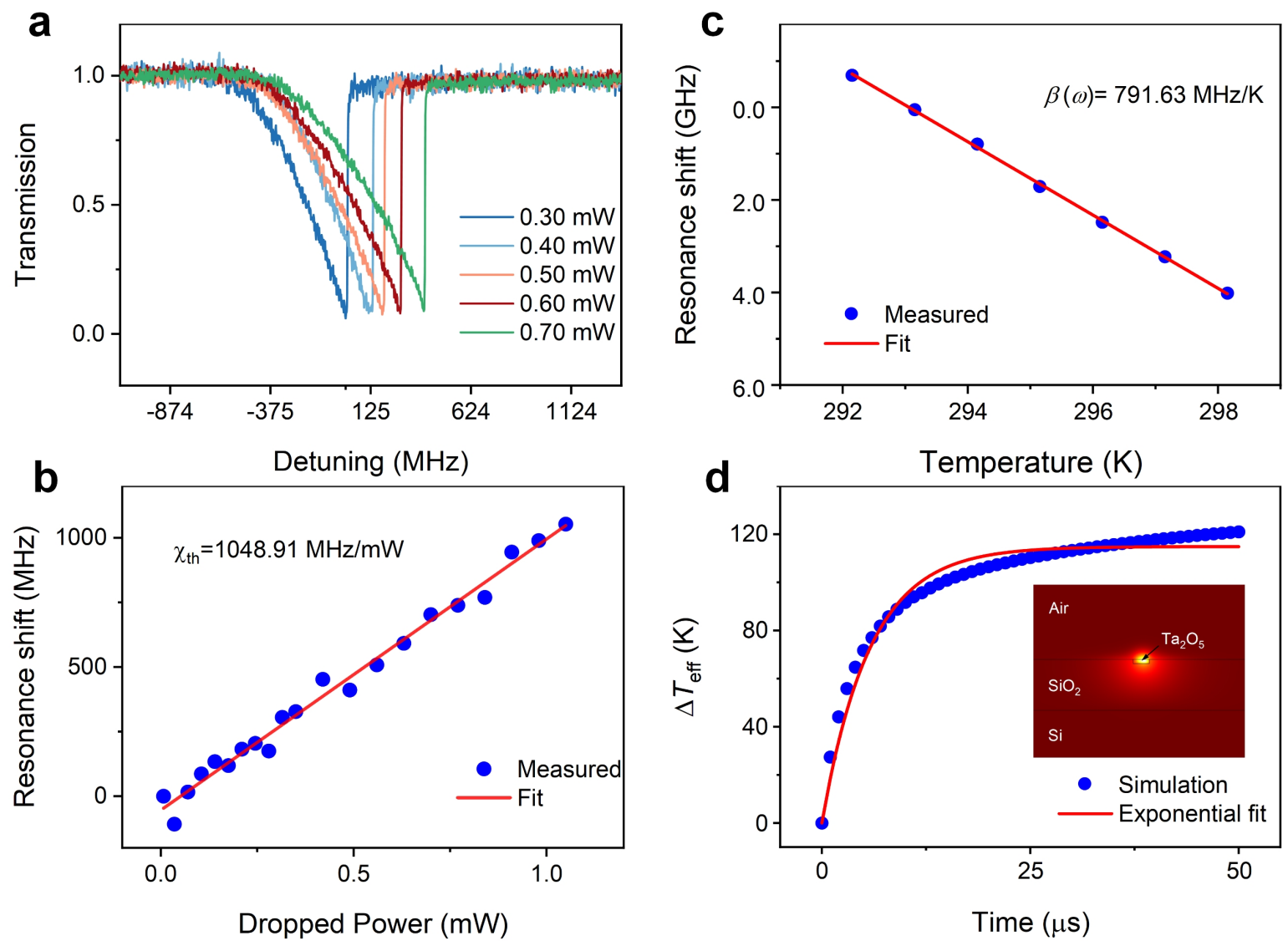

**Figure 3**. Thermo-optic effect analysis of the $Ta_2O_5$ MRR. **(a)** Transmission spectra at various on-chip input power showing an asymmetric resonance lineshape. **(b)** Resonance frequency shift ($\delta\omega$) as a function of dropped power ($P_d$), along with a linear fit for the extraction of $\chi_{th}$. **(c)** Resonance frequency shift as a function of temperature accompanied by a linear fit for determining $\beta(\omega)$. **(d)** Simulated $\Delta T_{eff}$ as a function of time, featuring an exponential fit to extract $K_C$. Inset: Simulated steady-state temperature distribution profile.

### *2.4 Dispersion engineering and generation of bright dissipative Kerr solitons in MRRs*

Leveraging the fabricated ultra-low-loss $Ta_2O_5$ waveguides and high $Q$ MRRs, we successfully demonstrated the generation of coherent Kerr soliton microcombs. Since anomalous dispersion is a fundamental prerequisite for soliton formation, we initially optimized the waveguide cross-sectional dimensions via FEM simulations, incorporating the measured material dispersion of the deposited $Ta_2O_5$ films (see Supplementary Note 9). To reliably excite the soliton state, we employed an experimental setup featuring an auxiliary laser for thermal stabilization, as illustrated in Figure 4(a). Prior to the soliton generation, the device dispersion was characterized utilizing a fiber Mach-Zehnder interferometer (MZI) assisted frequency scanning technique (detailed in Supplementary Note 9). The measured dispersion profile of the $Ta_2O_5$ MRR with a core cross-section of 1.6 μm (width) × 0.85 μm (height) is plotted in Figure 4(b). The extracted second-order dispersion parameter ($D_2/2\pi$) is approximately 1.17 MHz, confirming that the device operates well within the required anomalous dispersion regime. Subsequently, by tuning the pump laser frequency into the red-detuned regime of the MRR resonance using the setup shown in Figure 4(a), we recorded the spectral evolution of the intracavity optical field. As shown in Figure 4(c), the microcomb transitions from chaotic modulation instability (MI) states (MI-state I and MI-state II) to a bright single-soliton state, which exhibits the hallmark smooth $sech^2$ shaped spectral envelope.

Beyond the optical spectral characteristics, we further investigated the coherence evolution of the system via radio frequency (RF) beat note measurements. Because the repetition rate of the microcomb exceeds the bandwidth of conventional photodetectors, an electro-optic modulator was employed to generate an 88 GHz reference sideband from the pump laser. After filtering out the intense pump laser using a wavelength-division multiplexer (WDM), we heterodyned this reference sideband with the adjacent comb line. Figure 4(d) presents the corresponding RF spectra across three distinct evolutionary stages. During the two initial chaotic MI states, the absence of phase locking manifests as broadband amplitude noise in the beat note signal. Specifically, in MI-state I, the relatively concentrated energy among a limited number of comb lines results in high-amplitude noise accompanied by pronounced local fluctuations. As the spectrum further broadens into MI-state II, the finite pump energy redistributes across a larger number of newly excited comb lines. This spectral energy redistribution leads to a reduction in local optical power within the measurement band, thereby decreasing the overall RF noise intensity. Furthermore, the dense random interference smooths out the local fluctuations, yielding a flat noise envelope. Ultimately, as the system transitions into the single-soliton state, the robust phase locking among the comb lines significantly suppresses the broadband noise, presenting a sharp beat note peak with an exceptionally high signal-to-noise ratio (SNR). Considering the measured peak position at 20.32 GHz and the 88 GHz sideband offset, the repetition rate of the soliton is precisely determined to be 108.32 GHz, further corroborating its highly coherent and low-noise characteristics. From a device perspective, this 108.32 GHz repetition rate also places the present resonator in a lower repetition-rate regime than previous $Ta_2O_5$ bright-DKS demonstrations operating at several hundred gigahertz [15]. This operating regime is realized using the same tightly confined anomalous-dispersion device geometry in which multi-million $Q_i$ is obtained. The on-chip pump power used to access the single-soliton state was approximately 100 mW.

We next examined the experimentally accessible operating range of the single-soliton state under dual-pump thermal stabilization. For a qualitative platform-level reference, we characterized a stoichiometric $Si_3N_4$ microresonator with a $Q$ factor and free spectral range (FSR) comparable to those of the $Ta_2O_5$ device. Because the two platforms differ in their thermal responses and cavity characteristics, the main-pump and auxiliary-pump conditions were independently optimized for each device to access the single-soliton state. Under their respective optimized dual-pump conditions, the $Ta_2O_5$ resonator exhibited an experimentally accessible single-soliton window of approximately 7.5 GHz, whereas the corresponding window in the $Si_3N_4$ resonator was approximately 150 MHz [Fig. 4(e)].

The accessible soliton window in a dual-pump configuration is a system-dependent quantity whose width is influenced by multiple factors, including the cavity $Q$, coupling condition, optical absorption, pump power, scan conditions, auxiliary-laser parameters, mode crossings, and detuning calibration. The relatively weak thermo-optic response of $Ta_2O_5$ assists the reduction of thermally induced resonance shifts associated with intracavity-power changes during the transition to the soliton state, while the auxiliary laser provides additional thermal compensation. Under the operating conditions used here, these effects jointly favor a broader accessible single-soliton range. Accordingly, the present $Ta_2O_5$ resonator under the optimized dual-pump configuration exhibits a large pump-detuning tolerance, providing a wider operating margin for single-soliton access and reducing the sensitivity of soliton operation to precise pump-detuning control.

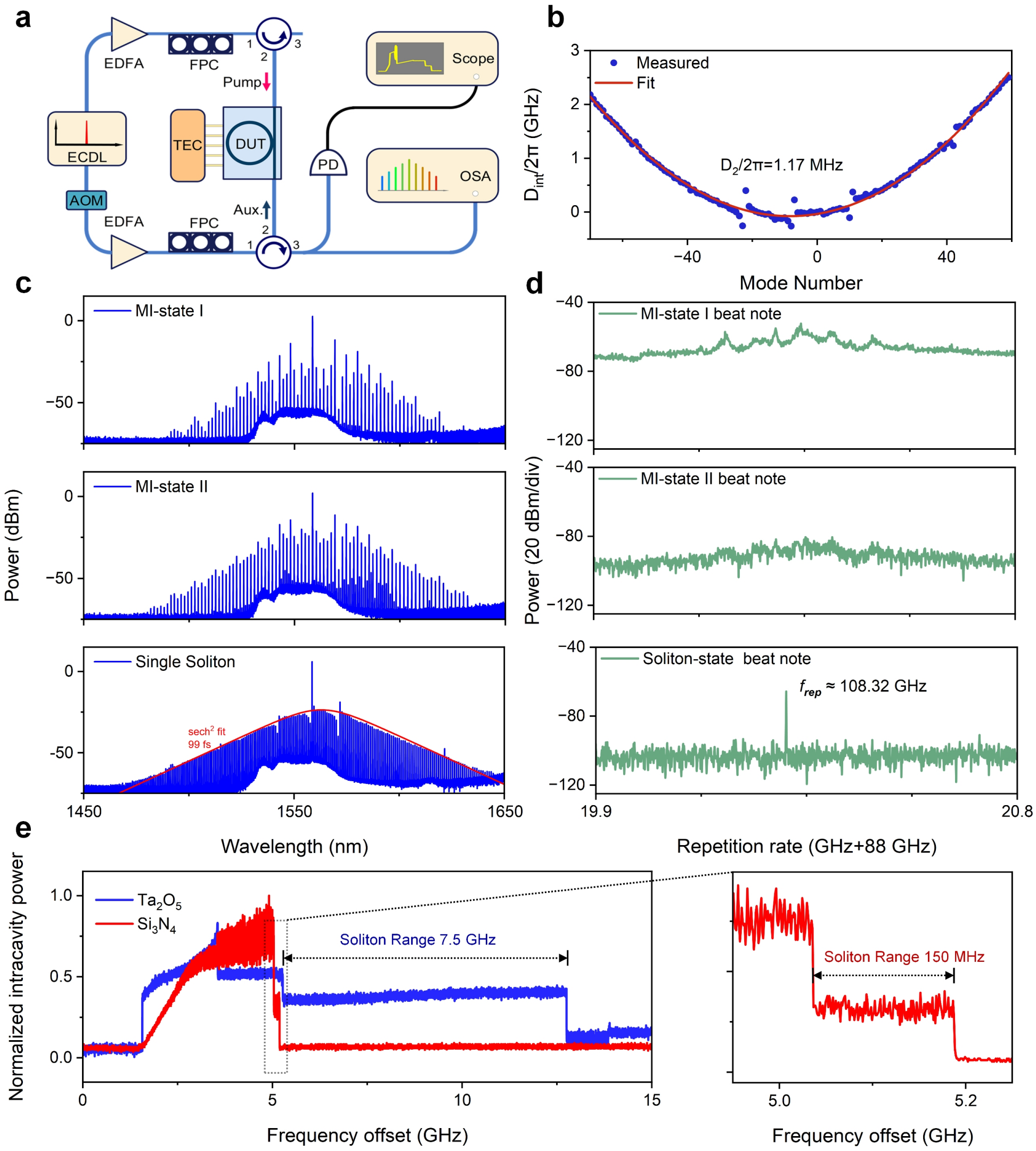

**Figure 4.** Generation and characterization of bright single-soliton microcombs in $Ta_2O_5$ MRRs. **(a)** Experimental setup. **(b)** Measured integrated dispersion with a parabolic fit (red curve) yielding $D_2/2\pi$ = 1.17 MHz. **(c)** Intracavity spectral evolution from chaotic MI states to a single-soliton state. The red $sech^2$ fit indicates a pulse duration of approximately 99 fs. **(d)** Corresponding RF beat note spectra. The sharp peak in the single-soliton state reveals a repetition rate ($f_{rep}$) of 108.32 GHz. **(e)** Normalized intracavity power traces of the $Ta_2O_5$ and $Si_3N_4$ MRRs under their respective optimized soliton-access conditions, showing experimentally accessible single-soliton windows of approximately 7.5 GHz and 150 MHz, respectively.

To evaluate the overall performance of the demonstrated $Ta_2O_5$ nonlinear photonic platform, Table 1 compares its material properties, device performance, waveguide geometry, and fabrication characteristics with those of other representative nonlinear photonic platforms used for Kerr microcomb generation. Compared with platforms such as $Si_3N_4$, AlN, and AlGaAs, $Ta_2O_5$ exhibits a relatively low thermo-optic coefficient, which is favorable for reducing thermally induced resonance shifts during nonlinear operation. Meanwhile, the experimentally inferred Kerr nonlinear refractive index of the present $Ta_2O_5$ platform is approximately (4.46 ~ 5.21) × $10^{-19}$ $m^2/W$ (Supplementary Note 11), approximately four times that of Hydex, which also exhibits a relatively low thermo-optic coefficient. Therefore, $Ta_2O_5$ provides a favorable balance between a relatively weak thermo-optic response and relatively strong Kerr nonlinearity, offering an advantageous material basis for high $Q$ nonlinear microresonators and soliton microcomb devices.

From a device and fabrication perspective, the fabrication flow employed in this work mitigates the trade-off between tight mode confinement and scattering loss. Sidewall roughness in conventionally etched $Ta_2O_5$ waveguides introduces additional scattering loss, while reducing the lateral waveguide dimensions and increasing mode confinement further enhance the interaction between the optical mode and the sidewalls, making the simultaneous realization of high $Q$ and dispersion engineering more challenging. Here, thermal reflow of the $SiO_2$ trenches reduces etched-sidewall irregularities, while CMP further decreases the top-surface roughness of the waveguides, thereby mitigating scattering loss. On this basis, although a lower-cost magnetron-sputtering process is employed compared with IBS, we still achieve $Q_i$ exceeding $4.0 \times 10^6$ in tightly confined 1.6 μm × 0.85 μm microring resonators satisfying the anomalous-dispersion requirement, corresponding to a propagation loss below 10 dB/m. This demonstrates a favorable balance among device performance, fabrication cost, and process scalability.

In addition, compared with thick-film LPCVD $Si_3N_4$ platforms, $Ta_2O_5$ offers greater process flexibility for thick-film deposition, alleviating fabrication constraints associated with high tensile stress and film cracking. This characteristic further facilitates the realization of thick-core $Ta_2O_5$ photonic devices with tight mode confinement and flexible dispersion engineering.

**Table 1. Comparison of performance of representative nonlinear material platforms for Kerr microcomb generation**

| Material | Film Deposition | Intrinsic $Q$ ($10^6$) | Nonlinearity $n_2$ ($\times 10^{-19}$ $m^2/W$) | Thermo-optic ($\times 10^{-5}$ $K^{-1}$) | Cross sections (μm × μm) |
|---|---|---|---|---|---|
| $Ta_2O_5$ (this work) | Sputter | > 4.0 | (4.46 ~ 5.21) | ~ 0.8 | 1.60 × 0.85 |
| $Si_3N_4$ [46] | LPCVD (> 700°C) | > 10 | ~ 2.4 | ~ 2.5 | 1.58 × 0.75 |
| $Ta_2O_5$ [15] | IBS | 3.8* | ~ 6.2 | ~ 0.8 | 5.00 × 0.57 |
| Hydex [11] | PECVD | ~ 1.5 | ~ 1.2 | ~ 1.0 | 1.45 × 1.50 |
| AlN [47] | Sputter/Epitaxy | ~ 3.0 | ~ 2.3 | ~ 2.0 | 3.50 × 1.00 |
| AlGaAs [45] | Epitaxy | ~ 2.0 | > 100 | > 20.0 | 0.40 × 0.70 |

*The maximum intrinsic quality factor $Q_i \approx 3.8 \times 10^6$ reported by Jung et al. was measured in a wider, approximately 5-μm-wide ring geometry. The same study demonstrated bright-DKSs in anomalous-dispersion resonators with cross sections of 1.60 × 0.57 $μm^2$ and 1.65 × 0.57 $μm^2$, for which the corresponding $Q_i$ values were not explicitly reported.

## 3. Conclusion

In conclusion, $Ta_2O_5$ combines a broad optical transparency window, relatively strong third-order nonlinearity, a weak thermo-optic response, and favorable thick-film growth characteristics, making it an attractive material platform for integrated Kerr nonlinear photonics. Building on these material advantages, this work addresses the challenge of jointly achieving tight mode confinement, low propagation loss, and anomalous dispersion in the $Ta_2O_5$ platform through systematic optimization of key fabrication processes, establishing an integrated $Ta_2O_5$ nonlinear photonic platform that combines high $Q$, tight mode confinement, and dispersion-engineering capability. In tightly confined $Ta_2O_5$ waveguides, the enhanced mode–sidewall interaction increases the sensitivity of the devices to sidewall roughness and scattering loss. Through $SiO_2$ trench reflow and CMP optimization, we achieve $Q_i$ exceeding $4.0 \times 10^6$ in 1.6 μm × 0.85 μm microring resonators, corresponding to propagation losses below 10 dB/m, while maintaining anomalous dispersion suitable for bright-DKS generation. To the best of our knowledge, this represents the highest $Q_i$ reported for tightly confined, thick-core $Ta_2O_5$ microring resonators. These results indicate that the fabrication flow effectively mitigates the trade-off between tight mode confinement and scattering loss in $Ta_2O_5$ waveguides.

Based on this platform, we further generate a coherent bright single-soliton microcomb with a repetition rate of 108.32 GHz, using an on-chip pump power of approximately 100 mW to access the single-soliton state. Under optimized dual-pump operating conditions, the $Ta_2O_5$ resonator exhibits an experimentally accessible single-soliton window of approximately 7.5 GHz, indicating a large pump-detuning tolerance. The relatively weak thermo-optic response of $Ta_2O_5$ helps reduce thermally induced resonance shifts associated with intracavity-power changes during transitions to the soliton state, while thermal compensation provided by the auxiliary laser further improves the accessibility of the single-soliton state.

Overall, this work demonstrates the joint realization of low propagation loss, tight mode confinement, anomalous dispersion, and bright-DKS operation in the $Ta_2O_5$ platform, while extending $Ta_2O_5$ bright-DKS operation toward the approximately 100 GHz repetition-rate regime. Meanwhile, the magnetron-sputtering process employed in this work offers a lower fabrication cost than IBS; together with $SiO_2$ trench reflow and CMP, it still enables high device performance, providing a favorable balance among device performance, fabrication cost, and process scalability. These results further highlight the potential of $Ta_2O_5$ for high $Q$, tightly confined integrated nonlinear photonic devices and provide a platform foundation for the development of more complex chip-scale nonlinear photonic systems.

**Funding.** Key Research and Development Program (2022YFA1404800), National Natural Science Foundation of China (12374359), National Natural Science Foundation of China (62205273).

**Acknowledgment.** The authors thank Dr. Chenyang Zhao, Dr. Jie Wang, and the Analytical & Testing Center of NPU for their assistance and for providing facilities in EBL, ICP, AFM, and SEM measurements.

**Disclosures.** The authors declare no conflicts of interest.

**Data Availability.** Data underlying the results presented in this paper are not publicly available at this time but may be obtained from the authors upon reasonable request.